\documentclass[smallextended]{svjour3}       
\usepackage{graphicx}
\usepackage{dcolumn}
\usepackage{bm}
\usepackage{graphicx}
\usepackage{latexsym,color}
\usepackage{amssymb}
\usepackage[colorlinks,citecolor=blue,urlcolor=blue,linkcolor=blue]{hyperref}

\begin{document}
\newcommand{\ti}[1]{\mbox{\tiny{#1}}}
\newcommand{\im}{\mathop{\mathrm{Im}}}
\def\be{\begin{equation}}
\def\ee{\end{equation}}
\def\bea{\begin{eqnarray}}
\def\eea{\end{eqnarray}}
\newcommand{\tb}[1]{\textbf{\texttt{#1}}}
\newcommand{\rtb}[1]{\textcolor[rgb]{1.00,0.00,0.00}{\tb{#1}}}
\newcommand{\btb}[1]{\textcolor[rgb]{0.00,0.00,1.00}{\tb{#1}}}
\newcommand{\il}{~}
\newcommand{\rc}{\rho_{\ti{C}}}
\newcommand{\dd}{\mathcal{D}}
\newcommand{\lie}{\mathcal{L}}
\newcommand{\Tem}{T^{\ti{(em)}}}
\newcommand{\Tm}{T^{\ti{M}}}
\newcommand{\Tkg}{T^{\ti{(KG)}}}
\newcommand{\g}[1]{\Gamma^{\phantom\ #1}}
\title{On the evolution equations for a  self-gravitating charged scalar field}
%
%
\author{Daniela Pugliese \and
        Juan A. Valiente Kroon}

\institute{Daniela Pugliese \at
              School of Mathematical Sciences, Queen Mary University of London\\
Mile End Road, London E1 4NS, UK\\
              \email{dpugliese@maths.qmul.ac.uk}           
           \and
           Juan A. Valiente Kroon \at
              School of Mathematical Sciences, Queen Mary University of London\\
Mile End Road, London E1 4NS, UK\\
               \email{j.a.valiente-kroon@qmul.ac.uk}           
               }%

\date{Received: date / Accepted: date}

\maketitle
\begin{abstract}
We consider   a complex scalar field minimally coupled to gravity and to a U(1) gauge symmetry and we construct of a first order symmetric
hyperbolic evolution system  for the Einstein-Maxwell-Klein-Gordon system.
Our analysis is based on a $1+3$ tetrad formalism which makes use of the
components of the Weyl tensor as one of the unknowns.
In order to ensure the symmetric hyperbolicity of the evolution equations, implied
by the Bianchi identity, we introduce
a tensor of rank 3 corresponding
to the covariant derivative of the Faraday tensor, and two tensors of rank 2 for the covariant derivative of the vector potential and the scalar field.
\end{abstract}

\section{Introduction}
In this article we discuss the construction of suitable
evolution equations for a self-gravitating
charged scalar field governed by the so-called
\emph{Einstein-Maxwell-Klein-Gordon system}.
It is well known that
General Relativity admits initial value problem
formulation whereby one prescribes certain initial data on a
3-dimensional hypersurface, and one purports to reconstruct the
spacetime associated to this initial data ---a so-called Cauchy
problem. The formulation of an initial value problem is a natural
starting point for a wide variety of analytical studies of the
qualitative properties of the solutions to the equations and  a
necessary starting point for the construction of the numerical
solutions. Examples of qualitative aspects of
the solutions requiring an suitable initial value formulation are the
discussion of local and global existence problems and the analysis of
the stability of certain reference solutions.

\medskip
In this article we approach the construction of the evolution
equations of the Einstein-Maxwell-Klein-Gordon system from the point
of view of mathematical Relativity. Hence, the amenability of our
analysis to analytic considerations takes precedence over numerical
considerations. Our discussion is
chiefly concerned with the construction of a first order system of
quasilinear hyperbolic evolution equations (\emph{FOSH}) out of the
Einstein-Maxwell-Klein-Gordon equations. In general, \emph{FOSH}
systems can be written in the following form:
\begin{equation}
\label{E:vesy}
\mathbf{A}^0\partial_t \textbf{{v}}-\mathbf{A}^j\partial_j \textbf{{v}}=\mathbf{B} \mathbf{v},
\end{equation}
where the variables of the system are collected in a $n$-dimensional
vector $\textbf{{v}}$, $t$ denotes a suitable time coordinate and
${\bm x}=(x^1,x^2,x^3)$ denotes some spatial coordinates and
$j=1,\,2,\,3$. The matrices $\mathbf{A}^0$ and $\mathbf{A}^j$ are
matrix valued functions depending on the coordinates $(t,{\bm x})$ and
the unknown $\textbf{{v}}$ ---that is, the system (\ref{E:vesy}) is,
in general, {\em quasilinear}. The system is said to be {\em symmetric
hyperbolic} if the matrices $\mathbf{A}^0$ and $\mathbf{A}^j$ are
symmetric and if $\mathbf{A}^0$ is a negative-definite matrix.

\medskip
As a consequence of the construction presented in this article, one
automatically obtains a local existence and uniqueness result for the
equations of the Einstein-Maxwell-Klein-Gordon system. The symmetric
hyperbolicity of the resulting evolution equations ensures that the
Cauchy problem for the systems is well-posed. In other words, if one
prescribes suitable initial data on an initial hypersurface, a unique
solution exists in a neighborhood of that hypersurface ---local in
time existence. The solutions depend continuously on the values of
initial data \cite{Reu98}. The question of the well-posedness of the evolution equations of the
Einstein-Maxwell-Klein-Gordon system is a problem that touches upon
many aspects of current theoretical and numerical analysis of many
physical phenomena. An suitable point of entry for the extensive
literature in this topic, with particular emphasis on numerics, is
given in \cite{LiePal12}. Scalar fields enter as important ingredients in
many theoretical models of contemporary physics. Here, we consider a
self gravitating scalar field, minimally coupled with the
gravitational field via the Einstein equations and the electromagnetic
field by the coupling constant $q$ ---the so-called {\em scalar boson
charge}. In the presence of strong gravitational fields, scalar fields
are described by the general-relativistic field equations. These
configurations arise in many areas of high-energy Cosmology as
inflationary models carried by scalar fields. Moreover, scalar fields
are also candidates for dark matter source. In astrophysics, scalar
fields arise in stellar models as configurations of hypothetical
scalar particles known as bosons stars, or as a boson core of very
compact stars \cite{topicalreview}.

\medskip
Our analysis is based on a $1+3$ tetrad formalism.
The hyperbolic reduction procedure described in the present article
borrows from the discussion of the evolution equations for the
Einstein-Euler system by H. Friedrich in
\cite{Fri98c} ---see also \cite{Fri91,Fri96,FriRen00} and the generalization to Einstein-Euler-Maxwell system in \cite{Pugliese:2011aa}. In this
reference a Lagrangian gauge was used to construct the required hyperbolic
reduction and to obtain the desired evolution equations. The central
equation in this discussion is the \emph{Bianchi identity}. It
provides evolution equations for the components of the Weyl
tensor. The addition of electromagnetic interactions to Friedrich's
system in \cite{Pugliese:2011aa} destroys, in principle, the symmetric hyperbolic nature of the
evolution equation as derivatives of the Faraday tensor enter into the
principal part of the Bianchi evolution equations. This difficult
was handled by the introduction of a new field unknown corresponding to
the derivative of the Faraday tensor for which suitable field and
evolution equations can be obtained. This strategy can be adapted to
the case of a  complex scalar field case coupled with the
electromagnetic field. In addition to the auxiliary variable
associated to the covariant derivative of the Faraday tensor, our
analysis required the use of two further tensors of rank 2 for the
covariant derivative of the vector potential and the scalar field. The
Lagrangian approach to the description of a scalar field follows the
ideas developed in \cite{AlhMenVal10} for a real scalar field. It is
important to observe that the extra gauge freedom induced by the frame
representation of the Einstein-Klein-Gordon system used in the present article
is associated to the evolution of the spatial frame coefficients along
the flow of the time-like frame ---in particular, in order to
implement the Lagrangian description it is  required that  the
timelike vector of the orthonormal frame follows the matter flow
lines. {In order to do this the scalar field  must satisfy  suitable  regularity conditions.
In fact, the local nature of the  treatment (fixing suitable  initial data  existence and uniqueness of a solution to the evolution equation can be established  locally
in time) assures  the scalar gradient to be timelike  only in a neighborhood of the initial data.  Afterwards this conditions may no longer be satisfied and the the scalar gradient can be  spacelike or null, and  in this case the entire set up breaks  down}. The remaining frame components are chosen to be Fermi
propagated along this direction.

\medskip
The present article is structured as follows: In
Section \ref{Sec:thebasiceq} we write and discuss the relativistic
equations describing a charged scalar field. The tetrad formalism
used in this article is briefly reviewed in
Section \ref{Sec:CHANGENOTATION}.  General remarks
concerning the reduction procedure to obtain suitable
evolution equations are given in Section \ref{Sec:redu}; the
resulting evolution  equations are discussed in
Section \ref{Sec:EB} and the subsequent sections. The auxiliary fields
and the gauge conditions are presented in Section \ref{Sec:NVGC}. A summary of the evolution equations is
given in Section \ref{Section:Summary}. Some concluding remarks are given in
Section \ref{Sec:Colcusione}.

\section{The charged scalar field equations}\label{Sec:thebasiceq}
In the present article we will consider the \emph{Einstein field equations}
\be
G_{\mu\nu}=\kappa T_{\mu\nu},
\ee
with matter source given by a self--gravitating
charged (i.e. complex) scalar field $\Phi$ minimally coupled to
gravity and to a $U(1)$ gauge field $A^{\mu}$. The total energy momentum
tensor for this system is given by
\be\label{E:aaa.0}
T_{\mu\nu}=T^{\ti{(KG)}}_{\mu\nu}+T^{\ti{(em)}}_{\mu\nu},
\ee
where, using units such that $\hbar=c=1$,
\be\label{E:aaa}
T^{\ti{(em)}}_{\mu\nu}\equiv
\frac{1}{4}g_{\mu\nu}F_{\lambda\rho}F^{\lambda\rho}-g^{\lambda\rho}F_{\mu\lambda}F_{\nu\rho},
\ee
denotes the energy momentum tensor of the \emph{free electromagnetic field} and
\bea\label{E:colonnasonora}
T^{\ti{(KG)}}_{\mu\nu}\equiv \left(D_{\mu}
\Phi\right)^{*}\left(D_{\nu}\Phi\right)+\left(D_{\mu}
\Phi\right)\left(D_{\nu}\Phi\right)^{*}-g_{\mu\nu}g^{\lambda\rho}\left(D_{\lambda}
\Phi\right)^{*}\left(D_{\rho}\Phi\right)
\eea
is the energy momentum tensor for the charged scalar field ---see
e.g. \cite{HawEll75}. In the previous equations
$D_\mu\equiv\nabla_\mu+\mbox{i} q A_\mu$, where the constant $q$ is
the \emph{boson charge} and $\nabla_\mu$ stands for the Levi-Civita covariant
derivative of the metric $g_{\mu\nu}$ of signature $(+,-,-,-)$. Moreover,
${}^*$ denotes the operation of complex conjugation while
\begin{equation}
F_{\mu\nu}\equiv \nabla_{\mu} A_{\nu}-\nabla_{\nu}A_{\mu}
\label{Definition:Faraday}
\end{equation}
is the electromagnetic field tensor (\emph{Faraday tensor}).

\medskip
In the sequel, it will be convenient to write the scalar field $\Phi$ in
terms of two real scalar fields $\theta$ and $\phi$ such that
\be
\Phi(x)=e^{i\theta(x)}\phi(x),\quad  \phi^*=\phi, \quad  \theta^*=\theta.
\ee
The energy momentum tensor of equation \il(\ref{E:colonnasonora}) then reads
\begin{eqnarray}
&& T^{\ti{(KG)}}_{\mu\nu}=\frac{\phi^2}{2}\bigg(\frac{2}{\phi^2}\nabla_\mu\phi\nabla_\nu\phi+
2\left(\sigma_\mu\sigma_\nu+q^2A_\mu
  A_\nu+2q\sigma_{(\mu}A_{\nu)}\right)\\
&& \hspace{2cm}
-g_{\mu\nu}\left(\frac{1}{\phi^2}g^{\lambda\rho}\nabla_\lambda\phi\nabla_\rho
  \phi+\sigma^2+q^2A^2+2q\sigma_\lambda A^\lambda\right)\bigg) \label{E:extendee}
,
\end{eqnarray}
where the vector field $\sigma_\mu$ is defined by
$\sigma_\mu\equiv\nabla_\mu\theta$, and we have used the notation
$A^2\equiv A_\lambda A^\lambda$. Consequently, it is possible to write
the tensor $\Tkg_{\mu\nu}$ as
\be\label{red}
\Tkg_{\mu\nu}=T^{(\phi)}_{\mu\nu}+\frac{1}{2}\phi^2T^{(\sigma)}_{\mu\nu}+\frac{1}{2}\phi^2 T^{\ti{(A)}}_{\mu\nu}+
\frac{1}{2}\phi^2 T^{(\sigma \ti{A})}_{\mu\nu},
\ee
where
\bea\label{E:puntophi}
&& T^{(\phi)}_{\mu\nu}\equiv \nabla_\mu\phi\nabla_\nu\phi
-\frac{1}{2}g_{\mu\nu}g^{\lambda\rho}\nabla_\lambda\phi\nabla_\rho\phi, \\
&& T^{(\sigma)}_{\mu\nu}\equiv
2\sigma_\mu\sigma_\nu -g_{\mu\nu}\sigma^2,
\\\label{E:puntosia}
&& T^{\ti{(A)}}_{\mu\nu}\equiv q^2(
2A_\mu A_\nu
-g_{\mu\nu}A^2), \\
&& T^{(\sigma\ti{A})}_{\mu\nu}\equiv 2q(
2\sigma_{(\mu}A_{\nu)}
-g_{\mu\nu}\sigma_\lambda A^\lambda).
\eea
Consistent with the requirement that $\nabla^\mu T_{\mu\nu}=0$, we require the scalar field $\Phi$ to satisfy the \emph{Klein-Gordon equation}
\[
D_\mu D^\mu \Phi=0.
\]
Equivalently, one has that
\be\label{E:tracephi}
\nabla_\mu\nabla^\mu\Phi+2i A^\mu\nabla_\mu\Phi+\mbox{i} g^{\mu\nu}\nabla_\mu A_\nu \Phi-A^2 \Phi=0.
\ee
\emph{In what follows, for simplicity of the presentation, we set
  $q=1$.}
\medskip

Written in terms of the real fields $\phi$ and $\sigma_\mu$, equation
(\ref{E:tracephi}) reads
\bea\label{E:forsifi}
&& \frac{1}{\phi}\nabla^\mu\nabla_\mu\phi -\sigma_\mu\sigma^\mu-A^2-2 A^\mu\sigma_\mu=0,
\\
\label{E:forsi}
&& \nabla^\mu\sigma_\mu+\frac{1}{\phi}\sigma_\mu\nabla^\mu\phi +\frac{1}{\phi}\sigma_\mu\nabla^\mu\phi+\frac{2}{\phi} A^\mu\nabla_\mu\phi+ g^{\mu\nu}(\nabla_\mu A_\nu )=0.
\eea

Finally, the electromagnetic field is described by the \emph{Maxwell equations} in the form
\be\label{E:Maom}
\nabla_{[\mu}F_{\nu\lambda]}=0,\qquad \nabla^\nu F_{\mu\nu}= \phi^2 (\sigma_\mu+A_\mu).
\ee

\section{Tetrad formalism}\label{Sec:CHANGENOTATION}

In the present article, the Einstein fields equations will be
expressed in terms of a frame formalism introduced in
\cite{Fri98c}. To this end, let $\{e_{a}{}\}_{a=0,...,3}$ denote a
basis of \emph{frame vectors} on the spacetime $(\mathcal{M},g_{\mu\nu})$
satisfying $g_{ab}\equiv g(e_{a}, e_{b})=\eta_{ab}=
\rm{diag}(1,-1,-1,-1)$. We denote by $\left(\omega^{a}{}\right)$ the
corresponding \emph{dual basis} (cobasis). Here, and in the rest of
the article, Latin letters $a, \,b,\ldots$ are used as spacetime frame
indices taking the values $0,\ldots,3$ while Greek letters
$\mu,\,\nu,\ldots$ denote the tensorial character of each object
---i.e. they are spacetime indices. The Latin letters
$i,\,j,\,k\,\ldots$ will be used as spatial frame indices taking the
values $1,\,2,\,3$.

\medskip
  The frame fields $e_a$ and the cobasis $\omega^a$
are expressed in terms of
a local coordinate basis as
\be\label{E:tetr1}
e_{a}=e_{a}{}^{\mu}\partial_\mu, \qquad \omega^a = \omega^a{}_\mu \mbox{d}x^\mu.
\ee
Thus,
\be\label{E:leggiadro1}
\omega^{a}_{\mu}e_{b}^{\mu}=
\delta_{b}{}^a,\qquad
\omega^{a}_{\mu}e_{a}^{\nu}=\delta_{\mu}{}^\nu,
\ee
so that the metric tensor can be written as
\be\label{E:applica100}
g_{\mu\nu}=\eta_{ab}\omega_{\mu}^{a}\omega_{\nu}^{b},\qquad g^{\mu\nu}=\eta^{ab}e^{\mu}_{a}e^{\nu}_{b}.
\ee
The \emph{commutation coefficients}  $D^{\phantom\ \phantom\ c}_{
ab}$  for the tetrad
$e_{a}^{\alpha}$ are defined by
\be\label{E:h2o1}
\left[e_{a},e_{b}\right]=
D^{\phantom\ \phantom\ c}_{
ab}e_{c}\equiv \left(e_{a}^{\mu}\partial_{\mu}
e_{b}^{\nu}-
e_{b}^{\mu}\partial_{\mu}e_{a}^{\nu}\right)e_{\nu}.
\ee
Finally, the \emph{connection coefficients} (or \emph{Ricci rotation
coefficients}) $\g{a}_{b\ c}$ for the tetrad $ e^{\alpha}_{a}$ are
defined by the relations
\be\label{E:trconn}
\nabla_{a}e_{b}=\g{c}_{a\ b}e_c,
\qquad\nabla_{a}\omega^{b}=-\g{b}_{a\ c}\omega^c.
\ee
Hence, equation (\ref{E:h2o1}) can be rewritten as
\be\label{E:h2o1xfile}
\left[e_{a},e_{b}\right]
=(\g{c}_{a\ b}-\g{c}_{b\ a})e_c.
\ee
In particular, since
$e_{a}\left(\eta_{bc}\right)=0$, one has that
\be\label{E:simmesnt}
\Gamma_{a(bc)}=0.
\ee
The components of the \emph{Riemann tensor} with respect to the frame $e_a$
are given in terms of the connection coefficients by
\be
\label{E:D41}
R^{a}_{\phantom \ {{bcd}}}=e_c(\g{a}_{ {{d\ b}}})-e_d(\g{a}_{ {{c\
b}}})-\g{a}_{e\ b}(\g{e}_{c\ d}-\g{e}_{d\ c})+\g{a}_{c\ e}\g{e}_{d\
b}-\g{a}_{d\ e}\g{e}_{c\ b}.
\ee
The components of the Riemann tensor admit the decomposition
\be\label{E:decom}
R_{abcd}=C_{abcd}+ \left\{g_{{a[c}}S_{{d]b}}-g_{{b[c}}S_{{d]a}}\right\},
\ee
where $S_{ab}$ denotes the \emph{Schouten tensor}
\be\label{E:S}
S_{ab}\equiv R_{ab}-\frac{1}{6}R g_{ab},
\ee
with $R_{ab}\equiv R^{c}_{\phantom\ {acb}}$ the components of the
\emph{Ricci tensor} and $R\equiv g^{ab}R_{ab}$ the \emph{Ricci
scalar}. Finally $C_{abcd}$ denotes the components of the \emph{Weyl
tensor} with respect to $e_a$. The components of the curvature tensor
satisfy the Bianchi identity
\be\label{E:CH3COOH}
\nabla_{[e}R^{a}_{\phantom\ bc]d}=0.
\ee
As it is well known, the contracted version of the above identity leads to
\begin{equation}\label{E:eqcr}
\nabla^{a}G_{ab}=0, \qquad G_{ab}\equiv R_{ab}-\frac{1}{2}R g_{ab},
\end{equation}
where $G_{ab}$ denotes the components of the \emph{Einstein tensor}. Defining the \emph{Friedrich tensor} $F_{abcd}$ via
\be\label{E:F1}
F_{abcd}\equiv C_{abcd}-g_{{a[c}}S_{{d]b}},
\ee
one concludes, furthermore, that
\be\label{E:hand}
F_{{abc}}\equiv  \nabla_{{d}}F^{{d}}{}_{{abc}}=0.
\ee
Taking the Hodge dual of equation (\ref{E:hand}) with respect to the index pair ${cd}$, we obtain another equation of the form  (\ref{E:hand}) for a tensor $\tilde{F}_{abcd}$ defined by
\be\label{E:F1STAR}
\tilde{F}_{abcd}\equiv C^*_{abcd}+\frac{1}{2}S_{{pb}}\epsilon^{{p}}{}_{\ {acd}},
\ee
where $\epsilon_{abcd}$ denotes the components of the \emph{completely
antisymmetric Levi Civita tensor} with respect to the frame $e_a$ and
$C^*_{abcd}\equiv \frac{1}{2}C_{{abef}}\epsilon^{{ef}}_{\phantom\
\phantom\ cd}$ ---see
e.g. \cite{Fri98c,BarMaaTsa07,EllEls98,MisThoWhe73}.

\section{The hyperbolic reduction procedure}\label{Sec:redu}
\subsection{General considerations  regarding  Friedrich's frame formulation of the Einstein field equations}\label{Sec:coondi}
Following \cite{Fri98c}, it is convenient to introduce here the
following notation:
\be
N^{a}\equiv\delta^{a}_0,\qquad N\equiv N^{a}e_a=e_0.
\ee
A tensor
$T_{a_1\cdots a_p}$ is said to be \emph{spatial} if any contraction
with with $N^a$ vanishes. The subspaces orthogonal to $N^a$ inherit
the metric $h_{ab} \equiv g_{ab} - N_a N_b$ (indices are raised and
lowered using $g_{ab}$). Thus, $h_a{}^b$ is the orthogonal projector
into these subspaces.

\medskip
For a given tensor, any contraction with $N$ will be denoted by
replacing the corresponding frame index by $N$. The projection of a
tensor with respect to $h_a{}^b$ will be indicated by ${}'$. Thus, a
tensor $T_{abc}$ one has that
\be
T'_{aNb}=h_{a}{}^{m}N^p h_{b}{}^q T_{mpq}.
\ee
A spatial vector satisfies $T_{a_1\cdots a_p} = T'_{a_1\cdots
a_p}$. In what follows, let $\epsilon_{abc}\equiv \epsilon'_{Nabc}$,
where $\epsilon_{0123}=1$. In terms of the latter one has the
decomposition
\be
\epsilon_{abcd}=2 \left(N_{[a}\epsilon_{b]cd}-\epsilon_{ab[c}N_{d]}\right).
\ee

Given a spatial tensor  we define the spatial covariant derivative
\be
\mathcal{D}_aT_{a_1\cdots a_p} \equiv \nabla_bT_{b_1 \cdots b_p}h_a{}^bh_{a_1}{}^{b_1}\cdots h_{a_p}{}^{a_p}.
\ee
In particular, it can be readily verified that
$\mathcal{D}_ah_{bc}=\dd_a\epsilon_{bcd}=0$.

In the sequel, it will be convenient to introduce the tensors
\be\label{E:dozero}
a^a=N^{b}\nabla_{b}N^a, \quad \chi_{ab}=h_{a}{}^c\nabla_c N_b, \quad  \chi\equiv h^{ab}\chi_{ab}.
\ee
These are related to the connection as can be see from
\be\label{E:rezero}
\nabla_aN^b= N_aa^b +\chi_a{}^b,\quad a^a=h_b{}^a\Gamma^b_{00},\quad\chi_{ab}=- h_a{}^c h_{b}{}^d\Gamma_{cd}^0.
\ee

\medskip
For future use, we notice the decomposition of the Weyl tensor as
\be\label{E:Cabcd}
C_{abcd}=2 (l_{b[c}\hat{E}_{d]a}-l_{a[c}\hat{E}_{d]b})-2(N_{[c}\hat{B}_{d]p}\epsilon^p_{\phantom\ ab}+N_{[a}\hat{B}_{b]p}\epsilon^p_{\phantom\ cd})
\ee
in terms of its electric $\hat{E}_{ab}=C'_{NaNb}$ and magnetic part
$\hat{B}_{ab}={C^{*}}'_{NaNb}$ relative to $N_a$, where $l_{ab}\equiv
h_{ab}- N_a N_b$. The Schouten tensor $S_{ab}$ will be expressed in
terms of the matter fields. Accordingly, we write
\be\label{E:sab}
S_{ab}=S^{\ti{(KG)}}_{ab}+S^{\ti{(em)}}_{ab},
\ee
where
\be\label{E:sab2}
S_{ab}^{\ti{(KG)}}=\kappa\big(T^{\ti{(KG)}}_{ab}-\frac{1}{3}g_{ab}g^{cd} T^{\ti{(KG)}}_{cd}\big),
\ee
with a similar expression for $S^{\ti{(em)}}_{ab}$. Using equations
\il(\ref{E:Cabcd})and \il(\ref{E:sab}) in equation \il(\ref{E:decom}),
the Riemann curvature tensor $R_{abcd}$, can be written in the form
\bea\label{E:recie}
R_{abcd}&=&R_{abcd}^{\ti{W}}+R_{abcd}^{\ti{KG}}+R_{abcd}^{\ti{em}},
\eea
where
\bea \nonumber R_{abcd}^{\ti{em}}&\equiv &
\kappa\bigg((E^2+B^2)\big(\frac{1}{2}\left(g_{a[c}g_{d]b}-g_{b[c}g_{d]a}\right)-2
N_{[b}g_{a][c}N_{d]}\big)\\
&& \hspace{0.3cm} -2E_{[b}g_{a][c}E_{d]}+2B_{[b}g_{a][c}B_{d]}+2
\big(N_{[a}g_{b][c}\mathcal{G}_{d]}-\mathcal{G}_{[a}g_{b][c}N_{d]}\big)\bigg),
\\ R_{abcd}^{\ti{W}}&\equiv&2
(l_{b[c}\hat{E}_{d]a}-l_{a[c}\hat{E}_{d]b})-2(N_{[c}\hat{B}_{d]p}\epsilon^p_{\phantom\
ab}+N_{[a}\hat{B}_{b]p}\epsilon^p_{\phantom\ cd}), \\
R_{abcd}^{\ti{KG}}&\equiv&g_{{a[c}}S^{\ti{KG}}_{{d]b}}-g_{{b[c}}S^{\ti{KG}}_{{d]a}},
\eea
with $S^{\ti{KG}}_{{db}}$ as defined in equation
\il(\ref{E:sab2}). Finally, from the Bianchi identity for the tensor
$F_{abcd}$, equation \il(\ref{E:hand}), we obtain the decomposition
\be
F_{abc}=N_a\left(F'_{NbN}N_c-F'_{NcN}N_b\right)-2 F'_{aN[b}N_{c]}+ N_{a}F'_{N bc}+F'_{abc},
\ee
where
\bea
\nonumber
 && F'_{aNb}=\mathcal{L}_{N}F'_{NaNb}+ \dd^cF'_{caNb}-a^c\left(F'_{Nacb}+F'_{caNb}\right)+ a_a F'_{NNNb}
\\
\label{E:decores}
&& \hspace{1cm} -\chi^{cd}F'_{cadb}-\chi_a^{\phantom\
c}F'_{NcNb}-\chi_b^{\phantom\ c}F'_{NaNc}+\chi^c_{\phantom\ a}
F'_{cNNb}+\chi F'_{NaNb}
\eea
with $\mathcal{L}_{N}$ denoting the Lie
derivative in the direction of $N^a$. From equation \il(\ref{E:F1}) we
compute the following components for the Friedrich tensor $F_{abcd}$
and $\tilde{F}_{abcd}$:
\bea
\tilde{F}'_{NNNa}&=&0, \qquad F'_{NNNa}=-\frac{ \kappa}{2} T_{ec}N^c h_a{}^e,
\\
\tilde{F}'_{NaNb}&=&\hat{B}_{ab}, \qquad F'_{NaNb}=\hat{E}_{ab}-\frac{\kappa}{2} T_{dc}h_b{}^d h_a{}^c,
\\
\tilde{F}'_{NNab}&=&\frac{\kappa}{2}\left(T_{pv}N^v\right)\epsilon{}^p_{uab}N^u,
\quad F'_{NNab}=0,
\\
\tilde{F}'_{aNNb}&=&-\hat{B}_{ab}+\frac{\kappa}{2}\left(T_{pv}N^v\right)\epsilon^p{}_{aub}N^u,
\\
F'_{aNNb}&=&-\hat{E}_{ab}+\frac{\kappa}{2}h_{ab}\left[T_{uv}N^vN^u\right],
\\
\tilde{F}'_{Nabc}&=& \hat{E}_{ap}\epsilon^p{}_{bc}+
\frac{\kappa}{2}\left(T_{pv}h_a{}^v\right)\epsilon^p{}_{ubc}N^u, \qquad F'_{Nabc}=- \hat{B}_{ap}\epsilon^p{}_{bc},
\\
\tilde{F}'_{aNbc}&=&- \hat{E}_{ap}\epsilon^p{}_{bc}+\frac{\kappa}{2}\left(T_{up}N^u\right)
\epsilon^p{}_{dvf}h_a{}^dh_b{}^v h_c{}^f,
\\
F'_{aNbc}&=&\hat{B}_{ap}\epsilon^p{}_{bc}-\frac{\kappa}{2}\left[h_{ab}\left(T_{fu}N^uh_c{}^f\right)-
h_{ac}\left(T_{fu}N^uh_b{}^f\right)\right],
\\
\tilde{F}'_{abNc}&=&-2 \hat{E}_{p[b}\epsilon^{\phantom\ p}_{a]\phantom\ c}+\frac{\kappa}{2}T_{pu}h_b{}^u\epsilon^p{}_{avc}N^v,
\\
F'_{abNc}&=&- \hat{B}_{cp}\epsilon^p{}_{ab}+\frac{\kappa}{2} h_{ac}T_{uv}N^uh^v_b,
\\
\tilde{F}'_{abcd}&=&-\hat{B}_{pq}\epsilon^p{}_{ab}\epsilon^q{}_{cd}+\frac{\kappa}{2}
\left(T_{pv}h_{[b}^v\right)\epsilon{}^p_{uef}h_{a]}{}^uh_c{}^eh_d{}^f,
\\
F'_{abcd}&=&2 \big(l_{b[c}\hat{E}_{d]a}-l_{a[c}\hat{E}_{d]b}\big)-\kappa
T_{uv}h_b{}^vh_{c[a}h_{d]}{}^u.
\eea

\subsection{Evolution equations for the electric and magnetic fields}\label{Sec:EB}
The Maxwell equations are given by
\be\label{E:MW}
\nabla_{[a}F_{b c]}=0,\quad\nabla^a F_{ab}=J_b,
\ee
where
\be\label{E:Maoms}
J_a\equiv- \phi^2 (\sigma_a+A_a).
\ee

As it is well known, the electromagnetic field tensor (Faraday tensor)
$F_{ab}$ can be split into its \emph{electric} and \emph{magnetic} parts, $E_{a}=F_{ab}N^{b}$ and  $B^{a}=\frac{1}{2}\epsilon^{abcd}N_{b}F_{cd}$,  components  with respects to a flow. More precisely, one has that
\be
F_{ab}=2 E_{[a}N_{b]} -\epsilon_{abcd}B^{c}N^{d},
\ee
where $N_aN^a=1$. Using the decomposition into electric and
magnetic parts, the electromagnetic energy-momentum tensor of equation \il(\ref{E:aaa}) can be written as
\be\label{E:ff2}
T^{(\ti{em})}_{a b}\equiv-\frac{1}{2} N_aN_b (E^2+B^2)+\frac{h_{ab}}{6}(E^2+B^2)+P_{ab}-2 \mathcal{G}_{(a}N_{b)},
\ee
where we have written $E^2\equiv E_aE^a$ and $B^2 \equiv B^aB_a$, and $P_{ab}$ denotes the symmetric, trace-free tensor given by
\be
P_{ab}\equiv\frac{h_{ab}}{3}(E^2+B^2)-(E_aE_b+B_aB_b),
\ee
and
\be
\mathcal{G}_{a}\equiv\epsilon_{auvd}E^uB^vN^d,
\ee
denotes the \emph{Poynting vector}. Projecting equations \il(\ref{E:MW}) along direction longitudinal and transverse to  the vector $N^b$,and  orthogonal to the  $N^b$,  one obtains the \emph{Maxwell evolution equations}
\bea\label{E:NE2a}
\dot{E}_{\langle f \rangle }&=&-2E^a h_{f[a}\nabla_{b]} N^b-\epsilon_{abcd}h_f{}^b\nabla^a(B^c N^d)-h_e{}^b J_b,
\\
\label{E:Nb2a}
\dot{B}_{\langle f \rangle}&=&-2B^a h_{f[a}\nabla_{b]} N^b+\epsilon_{abcd}h_f{}^b\nabla^a(E^cN^d),
\eea
and the \emph{Maxwell constraint equations}
\bea\label{E:Nb1a}
 D^aB_a&=&-\epsilon_{abcd}N^bE^c\nabla^a N^d,
\\
\label{E:NE1a}
D^aE_a&=&\epsilon_{abcd}N^a B^b\nabla^c N^d+N^bJ_b,
\eea
where $\dot{E}\equiv N^a\nabla_a E_b$ stands for the covariant time
derivative of $E_a$ along $N$, $\dot{w}_{\langle a \rangle}\equiv
h_a{}^b \dot{w}_b$ is the orthogonal projection of covariant time
derivatives along the $N^a$ and $D_aw_b\equiv h_a^uh_b^v\nabla_uw_v$
denotes the fully orthogonally projected covariant derivative of a
vector $w_a$. Finally, we note that
\[
\epsilon_{abcd}h_f{}^b\nabla^a(X^cN^d)=-\mbox{curl}
X_f+\epsilon_{afcd}N^a X^c \dot{N}^d,
\]
where $\mbox{curl} X_f\equiv\epsilon_{facd}N^d \nabla^a B^c$.

\subsection{Evolution equation for $\hat{B}_{ab}$}\label{Sec:EvWeu}

The evolution equation for the magnetic part of the Weyl tensor, $\hat{B}_{ab}$ is encoded in the component $\tilde{F}'_{(a|N|b)}$ of the Friedrich tensor. More precisely, one has that

\bea\nonumber
0= \tilde{F}'_{(a|N|b)}&=&\mathcal{L}_N \hat{B}_{ab}-D_d \hat{E}_{c ( a }\epsilon_{b)}{}^{dc}+2 a_c \epsilon^{cd}{}_{( a}\hat{E}_{b)d}-\chi^c_{\phantom\ (a}\hat{B}_{b) c}
\\\label{E:sa}
&&\hspace{0.5cm}-2 \chi_{(a}{}^{c}\hat{B}_{b) c}+\chi \hat{B}_{ab}- \chi_{cd}\hat{B}_{pq}\epsilon^{pc}{}_{\ (a}\epsilon^{dq}{}_{b)}+ \tilde{F}^{'(\ti{M})}_{(a|N|b)},
\eea
where $\tilde{F}^{'(\ti{M})}_{(a|N|b)}
\equiv\tilde{F}^{'(\ti{em})}_{(a|N|b)}+\tilde{F}^{'(\ti{KG})}_{(a|N|b)}$ can be written as
\bea
 && \tilde{F}^{'(\ti{M})}_{(a|N|b)}\equiv
 \frac{\kappa}{2}\mathcal{D}^c\big(T_{up}
 \epsilon^p{}_{cv(b}h_{a){}^u}N^v\big)-\frac{ \kappa}{2}
 \chi^{cd}\big(T_{pv}h_{(a}{}^v
 h_{b)}^f\big)\epsilon^p{}_{uef}h_{c}{}^uh_{d}^e \nonumber\\
&& \hspace{2cm}+\frac{\kappa}{2}\epsilon^p{}_{cu(b}\chi^c{}_{a)}N^u T_{pv}N^v.\label{E:1utri}
\eea
Substituting equations \il(\ref{E:aaa.0}) and \il(\ref{E:aaa}) in
equation \il(\ref{E:1utri}) we find that the electromagnetic contribution to
the evolution equation of $\hat{B}_{ab}$ is given by
\bea\nonumber  \tilde{F}^{'(\ti{em})}_{(a|N|b)}&\equiv&
-\frac{\kappa}{2}\bigg(\epsilon\mathcal{D}^c\big(N^v
\epsilon^p{}_{cv(b}(
h^u{}_{a)}F_{uq}F_{p}{}^{q}-\frac{1}{4}h_{a)p}F_{qs}F^{qs})\big)
\\ &&\hspace{1cm}-\chi^{ue}
\epsilon^p{}_{uef}(h^f{}_{(b}h^v{}_{a)}F_{vc}F_p^{\phantom\
c}-\frac{1}{4}h^f{}_{(b}h_{a) p}F_{qs}F^{qs})\nonumber \\
&& \hspace{1.5cm}-N^u
\epsilon^p{}_{cu(b}\chi^c{}_{a)}E_dF_{p}{}^{d}\bigg).
\eea
Notice that this last expression contains derivatives of the Faraday
tensor which cannot be replaced by means of the Maxwell
equations. These derivatives enter into the principal part of the
evolution equations and destroy the hyperbolicity of the evolution
equations for the magnetic part of the Weyl tensor. In order to deal
with this difficulty, in \cite{Pugliese:2011aa} an additional
variable, corresponding to the derivative of the Faraday tensor has
been introduced. This will be discussed in subsection \ref{Sec:psi}.

In order to obtain the contribution of the Klein-Gordon field to the
evolution equation of $\hat{B}_{ab}$, we use equation \il(\ref{red})
in equation \il(\ref{E:1utri}). The last two terms of equation
\il(\ref{E:1utri}) with $T_{ab}$ replaced by $T_{ab}^{\ti{KG}}$ are a
combination of the fields $\sigma_a$, $\phi$, $A_a$ and the derivative
$\nabla_a\phi$. Accordingly, we introduce the following field variable:
\be\label{E:intropsi}
\varphi_a\equiv \nabla_a\phi.
\ee
However, the term
\be
\label{E:term}
\frac{\kappa}{2}\mathcal{D}^c\left(\Tkg_{up} \epsilon^p{}_{cv(b}h_{a)}{}^uN^v\right),
\ee
in equation \il(\ref{E:1utri}), contains first and second derivatives
of the Klein-Gordon field. In order to recast the term
\il(\ref{E:term}) in a more convenient form, we first evaluate
$\nabla_c \Tkg_{ab}$.  Using equation \il(\ref{red}) one can write
\be\label{red1}
\nabla_c\Tkg_{ab}=\nabla_cT^{(\phi)}_{ab}+\frac{\phi^2}{2}\nabla_c(T^{(\sigma)}_{ab}+T^{\ti{(A)}}_{ab}
+T^{(\sigma \ti{A})}_{ab})+\phi \nabla_c \phi(T^{(\sigma)}_{ab}+T^{\ti{(A)}}_{ab}
+T^{(\sigma \ti{A})}_{ab}).
\ee
Important for our purposes is that the last term in the last equation, namely $\phi \nabla_c \phi(T^{(\sigma)}_{ab}+T^{\ti{(A)}}_{ab}
+T^{(\sigma \ti{A})}_{ab})$, contains only $\phi$ and $\varphi_a$. Now, consider a \emph{generic} energy momentum tensor
$$
T^{\ti{(X)}}_{ab}=c_{\ti{(X)}} \bigg(
X_{(a}Y_{b)}
-\frac{1}{2}g_{ab}g^{ef}X_e Y_f\bigg),
$$
 where $c_{\ti{(X)}}$ is a constant. A computation then shows that
 \bea
&& \nabla_c T^{\ti{(X)}}_{ab}=c_{\ti{(X)}} \bigg(
\frac{1}{2}(\nabla_c X_{a}Y_{b} +X_{a}\nabla_cY_{b}+\nabla_c
X_{b}Y_{a} +X_{b}\nabla_cY_{a})\nonumber \\
&& \hspace{3cm}-\frac{1}{2}g_{ab}g^{ef}\nabla_c X_e Y_f-\frac{1}{2}g_{ab}g^{ef} X_e
\nabla_cY_f\bigg). \label{II}
 \eea
Applying this last formula to the energy-momentum tensors in \il(\ref{E:puntophi})-(\ref{E:puntosia}) one finds that
\bea
\nabla_c T^{(\sigma)}_{ab}&=&c_{(\sigma)} \left(
( \sigma_{a}\nabla_c\sigma_{b} +\sigma_{b}\nabla_c\sigma_{a})
-g_{ab}g^{ef} \sigma_e \nabla_c\sigma_f\right),
\\\
\nabla_c T^{\ti{(A)}}_{ab}&=&c_{\ti{(A)}} \left(
( A_{a}\nabla_cA_{b} +A_{b}\nabla_cA_{a})
-g_{ab}g^{ef} A_e \nabla_cA_f\right),
\\\label{LO}
\nabla_c T^{(\phi)}_{ab}&=&c_{(\phi)} \left(
(\phi_{a} \nabla_{c}\phi_b +\phi_{b}\nabla_c\phi_{a})
-g_{ab}g^{ef} \phi_e \nabla_c\phi_f\right),
 \eea
and, moreover, that
\bea
&& \nabla_c T^{(\sigma \ti{A})}_{ab}=c_{(\sigma \ti{A})} \bigg(
\frac{1}{2}(A_{b}\nabla_c \sigma_{a} +A_{a}\nabla_c \sigma_{b}
+\sigma_{a}\nabla_cA_{b} +\sigma_{b}\nabla_cA_{a}) \nonumber \\
&& \hspace{3cm} -\frac{1}{2}g_{ab}g^{ef}(A_f\nabla_c \sigma_e + \sigma_e
\nabla_cA_f)\bigg). \label{IIO}
\eea
In order to write this last expression in a form suitable to our purposes we introduce the  following two auxiliary fields:
\be\label{E:siai}
\psi_{ab}\equiv\nabla_a\sigma_b, \qquad
\zeta_{ab}\equiv\nabla_aA_b.
\ee
By introducing these new variables we remove all explicit derivatives
from the term (\ref{E:term}). Of course, the price paid by this is
that we have to find suitable evolution equations for the new
auxiliary fields.

\subsection{Evolution equation for $\hat{E}_{ab}$}
The evolution equation for the electric part of the Weyl tensor can be
obtained by expanding the components $F'_{(a|N|b)}-\frac{1}{2}h_{ab}h^{uv}F'_{uNv}$ of the Friedrich tensor. A lengthy computation shows that
\bea\nonumber
&& 0= F'_{(a|N|b)}-\frac{1}{2}h_{ab}h^{uv}F'_{uNv}\\
&&\phantom{0}=\mathcal{L}_N \hat{E}_{ab}+D_c\hat{B}_{d(a}\epsilon_{b)}{}^{cd}-2  a_c \epsilon^{cd}{}_{(a}\hat{B}_{b)  d}
-3 \chi_{(a}^{\phantom\ \phantom\ c}\hat{E}_{b)
  c}-2\chi^{c}_{\phantom\ (a}\hat{E}_{b) c}
\nonumber \\
&& \hspace{2cm}+h_{ab}\chi^{cd} \hat{E}_{cd}+2\chi \hat{E}_{ab}+
\bigg(F_{(a|N|b)}^{'(\ti{M})}-\frac{1}{2}h_{ab}h^{uv}F^{'(\ti{M})}_{uNv}\bigg), \label{E:terete}
\eea
where the matter contribution is given by
\bea \nonumber
&&F^{'(\ti{M})}_{(a|N|b)}-\frac{1}{2}h_{ab}h^{uv}F^{'(\ti{M})}_{uNv}
\nonumber \\
&& \hspace{2cm}=\frac{1}{4} \kappa \mathcal{L}_N \bigg(T_{dc}h_{a}{}^d
h_{b}^c\bigg)+\frac{1}{2}\kappa\bigg(h_{(a}{}^v\mathcal{D}_{b)}-\frac{1}{2}h_{ab}\mathcal{D}^v\bigg)(T_{uv}N^u)
\nonumber
\\
\nonumber && \hspace{2cm}+\kappa T_{uv}N^v\bigg(\frac{1}{2}h_{ab}a^u-h_{(b}{}^u
a_{a)}\bigg) +\frac{1}{2}\kappa T_{uv}N^uN^v
\bigg(\chi_{(ab)}-\frac{h_{ab}}{2}\chi\bigg) \\
&& \hspace{2cm}+\frac{1}{2}\kappa
\bigg(2\chi_{(a}{}^{u}h_{b)}{}^v-\chi_{\phantom\
(a}^uh_{b)}{}^v-\frac{1}{2}\chi^{uv}h_{ab}\bigg)
T_{uv}.
\label{E:forE}
\eea
In the case of the electromagnetic field
the above expression yields explicitly that
\bea \nonumber
&&F^{'(\rm{em})}_{(a|N|b)}-\frac{h_{ab}h^{uv}}{2}F^{'(\rm{em})}_{uNv}
\nonumber \\
&&=-\frac{1}{4}\kappa\mathcal{L}_N\left(h^q{}_{(a}h^c{}_{b)}F_{qf}F_c{}^{f}-
\frac{1}{2}h_{ab}F_{qp}F^{qp}\right)
-\frac{1}{2}\kappa\left(\frac{h_{ab}\mathcal{D}^v}{2}-h^v{}_{(a}\mathcal{D}_{b)}\right)
E^c F_{vc} \nonumber  \\
&&-\kappa E^c
F_{vc}\left(h^v{}_{(a}a_{b)}-\frac{1}{2}h_{ab}a^v\right)-
\frac{1}{2}\kappa F_{ud}F_{v}{}^{d}\left(2\chi_{(a}{}^{v}h_{b)}{}^u-
\chi^{v}{}_{(a}h_{b)}{}^u-\frac{1}{2}\chi^{uv}h_{ab}\right) \nonumber
\\
&&+\frac{1}{4}\kappa \chi_{(ab)}F_{qp}F^{qp}-\frac{1}{2}\kappa E^2
\chi_{(ab)}-\frac{1}{8}\chi h_{ab}F_{qp}F^{qp}+\frac{1}{4}
\kappa \chi h_{ab}E^2.
\eea
 As in the previous subsection we observe
the presence of derivatives of the Faraday tensor which need to be
dealt with by the introduction of a new field if one is to preserve
the hyperbolicity of the equations.

Finally, using equation (\ref{red}) we obtain the expression for
the scalar field contribution. In this case the first two terms of the
general expression (\ref{E:forE}), namely,
\be
\frac{1}{4} \kappa \mathcal{L}_N \left(\Tkg_{dc}h^d_{a}h^c_{b}\right)+\frac{1}{2} \left(h^{v}_{(a}\mathcal{D}_{b)}-\frac{1}{2}h_{ab}\mathcal{D}^v\right)\Tkg_{uv}N^u,
\ee
clearly contain derivatives of the Klein-Gordon energy-momentum
tensor. These terms can be rewritten using equation (\ref{red1}) and
introducing the auxiliaries fields of equations (\ref{E:intropsi})
and (\ref{E:siai}).

\subsection{Evolution equations for the auxiliary field $\psi_{abc}$}\label{Sec:psi}

The analysis of the evolution equations for the electric and magnetic
parts of the Weyl tensor led us to introduce the covariant derivative
of the Faraday tensor as further field variable. Accordingly, we set
\be
\label{E:tono}
\psi_{abc} \equiv \nabla_a F_{bc}.
\ee
It has been shown in \cite{Pugliese:2011aa} that, applying a covariant
derivative to the Maxwell equations (\ref{E:MW}), commuting covariant
derivatives and using the definition (\ref{E:tono}) one obtains the
following equations for the tensor $\psi_{abc}$
\be\label{E:star}
\nabla^b\psi_{adb}=2F^{eb}R_{[d|ae|b]}- \nabla_a J_d,
\qquad \label{E:stars}
\nabla^a {}^*\psi_{cab}=\epsilon_b^{\phantom\ aud}F_{eu}R^e_{\phantom\ dac},
\ee
where
\be\label{E:nemor}
{}^*\psi_{abc} \equiv \frac{1}{2} \epsilon_{bc}{}^{ef} \psi_{aef}.
\ee
As $\psi_{abc}=\psi_{a[bc]}$, one can naturally define its \emph{electric} and \emph{magnetic parts} respect to $N_a$ as
\bea
\mathcal{E}_{ad}\equiv\psi_{adn}N^n,
\qquad
\mathcal{B}_{au}\equiv\frac{1}{2}\epsilon_u{}^{vzt}N_v \psi_{azt}.
\eea
Notice that by construction one readily has that
$\mathcal{E}_{a}N^a=\mathcal{B}_aN^a=0$. Projecting the equations in
\il(\ref{E:star}) along the directions longitudinal and transverse to
$N^b$ one obtains the following set of evolution equations:
\bea\label{E:NE2ab}
\dot{\mathcal{E}}_{e\langle f \rangle}&=&-2\mathcal{E}_e{}^a h_{f[a}\nabla_{b]} N^b-\epsilon_{abcd}h_f{}^b\nabla^a(\mathcal{B}_e{}^c N^d)+ h_f{}^d\mathcal{S}_{ed},
\\
\label{E:Nb2ab}
\dot{\mathcal{B}}_{e \langle f \rangle }&=&-2\mathcal{B}_e{}^a h_{f[a}\nabla_{b]} N^b+\epsilon_{abcd}h_f{}^b\nabla^a(\mathcal{E}_e{}^cN^d)- h_f{}^d\mathcal{V}_{ed}.
\eea
It is noticed that the corresponding constraint equations assume the form
\bea\label{E:Nb1ab}
 D^a\mathcal{B}_{ae}&=&-\epsilon_{abcd}N^b\mathcal{E}_e{}^c\nabla^a N^d+ N^d\mathcal{V}_{ed},
\\
\label{E:NE1ab}
 D^a\mathcal{E}_{ea}&=&\epsilon_{abcd}N^a \mathcal{B}e{}^b\nabla^c N^d- N^d\mathcal{S}_{ed},
\eea
where for ease of presentation we have set
\bea
\mathcal{S}_{fd}\equiv 2F^{eb}R_{[d|fe|b]}-\nabla_f J_d,
\quad
\mathcal{V}_{fb}\equiv-\epsilon_f^{\phantom\ aud}F_{eu}R^e_{\phantom\ dab}.
\eea

\subsection{Remarks concerning the unknowns and gauge conditions}\label{Sec:NVGC}
For convenience, we collect the unknowns discussed in the previous sections in the following vector variable:
\be\label{E:v}
\mathbf{v}=\{e_a{}^\mu,\g{a}_{b\ c},\hat{E}_{ab},\hat{B}_{ab},A_{a},\Phi,E_a,B_a,\mathcal{E}_a,\mathcal{B}_a\}.
\ee
As already discussed, the matter model under consideration makes
further use of field $A_a$ and the phase of $\Phi=\phi e^{i
\theta}$. Accordingly one has to introduce the following auxiliary
fields:
\be
\sigma_a\equiv\nabla_a\theta,\qquad \psi_{ab}\equiv\nabla_a\sigma_b,\qquad\zeta_{ab}\equiv\nabla_aA_b,
\ee
{
Recall also, that in equation (\ref{E:intropsi}) we have also
introduced the 1-form $\varphi_a\equiv\nabla_a\phi$. Following the
discussion in \cite{AlhMenVal10} we use the real scalar field $\phi$ to
construct a frame adapted to the problem. Accordingly, we assume now
that $\varphi_a$ is timelike and consider a frame satisfying}
\be\label{E:realt}
\varphi_a=\alpha e_0,\qquad |\alpha|\equiv|\varphi|.
\ee
{In fact, in order to construct the adapted frame  for this problem, the gauge choice has been  based on some specific regularity assumptions on the scalar field, requiring the timelike vector of the orthonormal
frame to follow the matter flow.
Firstly, it is assumed  that
$\phi\in \mathcal{C}^{\infty}(\mathcal{M})$ and,
 secondly
we assume   $\nabla_a\phi$
is  and remains timelike. In fact, if the gradient is null or spacelike the system evolution  breaks down.}
Hence, one can write $\varphi^a=\alpha \delta_0^{a}$. It follows then that
\be
D_i\phi=0,\qquad {e'}_i^{\mu}\nabla_{\mu}\phi=0.
\ee
In terms of components respect to a coordinate basis one finds that
\be
e_0^{\mu}=\frac{\nabla^{\mu}\phi}{\alpha}.
\ee

\subsection{Evolution equation for the fields $\phi$ and $\varphi_a$}
The evolution equation for the real scalar field $\phi$ can be
recovered from the definition of $\varphi_a$. Namely, one has that
\be
\label{E:fie} \partial_t\phi=\varphi_t.
\ee
An evolution equation
for $\varphi_a$ can be found by using equation
\il(\ref{E:forsifi}). One has then that
\be
\label{E:forsifia}
\nabla^a\varphi_a=\phi(\sigma_a\sigma^a+A^2+2
A^a\sigma_a).
\ee
In view of the gauge condition in equation
\il(\ref{E:realt}), equation \il(\ref{E:forsifia}) is, in fact, an
equation for $\alpha$ and the connection coefficients of the
frame. The corresponding equations will be discussed in details in
Section \il(\ref{Sec:terean}).

\subsection{The evolution equation for   the vector potential $A_a$}
The evolution equation for the vector potential $A_a$ can be readily
obtained from the expression of the Faraday tensor in terms of the
curl of the vector potential ---cf. equation
(\ref{Definition:Faraday}). This equation is almost in hyperbolic
form. In order to close the system one has to introduce a
\emph{gauge source function} ---see e.g. \cite{Fri91}. This feature is
closely related to the gauge freedom built into equation
(\ref{Definition:Faraday}) and allows to specify freely the divergence of $A_a$. One has that
\bea\label{E:Aeq}
\nabla_0 A_b=F_{0b}+\nabla_b A_0,
\\ \label{E:Aeqt}
\nabla^0 A_0=\zeta(x)-\nabla^i A_i,
\eea
where $\zeta(x)\equiv \nabla_a A^a$. The evolution equations for
$F_{ab}$ have already been discussed in Section \il\ref{Sec:EB}.

\subsection{The evolution  equation for $\sigma_a$}
A key observation for our purposes is that $\theta$, the phase of the
complex scalar field $\Phi$, does not appear explicitly in the
evolution equations hitherto discussed, appearing only through its
derivative. These derivatives are, in turn, replaced by
$\sigma_a\equiv\nabla_{a}\theta$. The evolution equation for the
vector potential $\sigma_a$ can be inferred from the following
integrability condition
\be
\label{No-Torsion-co}
\nabla_a \sigma_b-\nabla_b \sigma_a=0,
\ee as
the connection being considered is torsion-free. The structure of this
equation is similar to that of equation (\ref{Definition:Faraday}). In
order to bring it to hyperbolic form it has to be complemented with
information about the divergence of $\sigma_a$ prescribed by a gauge
source function. From equation \il(\ref{No-Torsion-co}) if follows
that
\bea\label{E:sigma}
&& \nabla_0 \sigma_b=\nabla_b \sigma_0,
\\\label{E:sigmat}
&& \nabla^0 \sigma_0=\psi(x)-\nabla^i \sigma_i,
\eea
where $\psi\equiv\nabla^a\sigma_a $. The gauge source function $\psi$
is not truly independent. In fact, as a consequence of the Klein Gordon
equation (\ref{E:forsi}) one finds that
\be
\label{E:forsinowma}
\psi+ {\zeta}=-\frac{2\alpha}{\phi}\left(\sigma^0 +A^0\right)e_0.
\ee
Thus, one readily sees that $\psi$ can be expressed in terms of
$\zeta$ and further field variables.

\subsection{The  evolution equations for the auxiliary field $\psi_{ab}$ and  $\zeta_{ab}$}\label{Sec:gauge}
In order to motivate the construction of suitable evolution equations
for the auxiliary fields $\psi_{ab}$ and $\zeta_{ab}$, we consider
first the commutator of covariant derivatives acting on a generic
tensor $X_a$. One has that
\be\label{E:lstc}
\nabla_{c}\nabla_{d}X_{b}-\nabla_{d}\nabla_{c}X_{b}=R^{a}_{\phantom\ bcd}X_{a}
.
\ee
Making use of the definition of $Y_{bc}\equiv\nabla_bX_c$ we rewrite
equation \il(\ref{E:lstc}) as
\be\label{E:in}
\nabla_{c}Y_{db}-\nabla_{d}Y_{cb}=R^{a}_{\phantom\ bcd}X_{a},
\ee
for the tensor $Y_{bc}$. making use of this equation for
$X_a=\sigma_a,\,A_a$ one obtains
\bea\label{E:ina}
\nabla_{c}\psi_{db}-\nabla_{d}\psi_{cb}=R^{a}_{\phantom\ bcd}\sigma_a,
\\
\label{E:inb}
\nabla_{c}\zeta_{db}-\nabla_{d}\zeta_{cb}=R^{a}_{\phantom\ bcd}A_a,
\eea were the components of the tensor $R^a{}_{bcd}$ in the right-hand
side of equations (\ref{E:ina}) and (\ref{E:inb}) can be reexpressed,
using equation (\ref{E:decom}), as an algebraic expression involving
the Weyl tensor and the matter fields ---cf. equations
(\ref{E:sab})-(\ref{E:sab2}).  Thus, the right-hand side is a function
of $\mathbf{v}$ only. The antisymmetry with respect to the covariant
derivatives in the left-hand side of equations (\ref{E:ina}) and
(\ref{E:inb}) suggests that they imply symmetric hyperbolic evolution
equations if information concerning the divergence of fields
$\psi_{ab}$ and $\zeta_{ab}$ is provided. This is done in the next
sections.

\subsubsection{Evolution equation for the field $\zeta_{ab}$}\label{E:zetafor}

For convenience, define the tensor $g_a\equiv\nabla^c
\zeta_{ca}$. If $g_a$ is explicitly known in terms of other field
variables, then suitable hyperbolic equations for $\zeta_{ab}$ are
given by
\begin{eqnarray}\label{E:stasst}
&& \nabla_0 \zeta_{ia}- \nabla_i \zeta_{0a} = R^e{}_{0ia}A_e,
\\
\label{E:stasstt}
&& \nabla^0 \zeta_{0a}+ \nabla^i \zeta_{ia} = g_a(x),
\end{eqnarray}
The required information about $g_a$ can be deduced from the Maxwell equation
\be
\nabla^aF_{ab}=\nabla^a\nabla_a A_b-\nabla^a\nabla_b A_a=-\phi^2 (\sigma_b+A_b).
\ee
Commuting covariant derivatives one obtains from the latter that
or
\be\label{E:EAemi}
\nabla^a\zeta_{ab}- \nabla_b \nabla^a A_a -R^{ca}{}_{ab}A_c=-\phi^2 (\sigma_b+A_b).
\ee
Finally, noticing that $\zeta(x)=\nabla^a A_a$ and $\nabla^a \nabla_a A_b=\nabla^a\zeta_{ab}=g_b$ one obtains
\[
g_b= \nabla_b \zeta + R^{ca}{}_{ab}A_c-\phi^2 (\sigma_b+A_b).
\]
Thus, the divergence $\nabla^c \zeta_{ca}$ can be rewritten in terms
of the derivatives of the gauge source function (which are explicitly
known) and an algebraic expression of matter fields.

\subsubsection{Evolution equation for the field $\psi_{ab}$}\label{Sec:forpsia}

The analysis of the equation \il(\ref{E:ina}) is similar to that of
equation (\ref{E:inb}). Assuming that the divergence
$h_a\equiv\nabla^c \psi_{ac}$ is known, one readily obtains the
symmetric hyperbolic system
\bea
\label{E:psi}
&& \nabla_0 \psi_{da}- \nabla_d \psi_{0a} = R^e{}_{a0d} \sigma_e,
\\
\label{E:psit}
&& \nabla_0 \psi_{0a}+ \nabla_i \psi_{ia} = h_a.
\eea
Now, exploiting that $\psi_{ab}=\psi_{(ab)}$, one has that
\[
\nabla^a \psi_{ab} = \nabla^a \psi_{ba} = \nabla^a \nabla_b \sigma_a = \nabla_b \nabla^a \sigma_a + R^{ea}{}_{ab}\sigma_e.
\]
Hence, one finds that
\[
h_b = \nabla_b \psi  + R^{ea}{}_{ab}\sigma_e.
\]
The derivative $\nabla_b \psi$ can be computed, in turn, using
equation \il(\ref{E:forsinowma}). Hence, one finds that $h_b$ can be
expressed in terms of the derivatives of the gauge function $\zeta$
and auxiliary fields.

\subsection{Evolution equations for frame coefficients and connection coefficients}\label{Sec:terean}
In order to construct evolution equations for the tetrad and connection coefficients we follow closely the procedure indicated in \cite{AlhMenVal10}.

First of all, consider equation \il(\ref{E:h2o1xfile}) with
$N=e_0=\alpha\mathbf{\varphi}$ and $\varphi^{\mu}=\alpha e_0^{\mu}$,
$\varphi_{\mu}= e^0_{\mu}/\alpha$. In other words, we require the
timelike vector of the orthonormal frame to follow the \emph{``matter
flow lines $\nabla\phi$''}, ---see Section
\il\ref{Sec:gauge}{ and Section\il\ref{Sec:NVGC}}. Moreover, we introduce a coordinate system
$(x^\mu)$ by requiring the \emph{Lagrange condition}
$N^{\mu}=e^{\mu}_0=\delta^{\mu}_0$ ---see Section
\il\ref{Sec:coondi}). Taking into account the latter and using
equation \il(\ref{E:h2o1}) we find that
\be\label{E:h2o1oboe}
\partial_t e^{\mu}_i=(\Gamma_{0\ i}^{\phantom\ j }-\Gamma_{i\ 0}^{\phantom\ j })e^{\mu}_j+\Gamma_{0\ i}^{\phantom\ 0 }e^{\mu}_0.
\ee
The later is the required evolution equation for $e^{\mu}_i$.

We turn now the attention to the evolution equation for the connection
coefficients $\Gamma_{a\ b}^{\phantom\ c }$. The requirement that the
frame components $e^{\mu}_i$ are Fermi propagated along $e_0$ readily
implies that $\Gamma_{0\ i}^{\phantom\ j }=0$ ---see
e.g. \cite{Fri96}. Thus, no evolution equations are required for these
components. The only components requiring evolution equations are
$\Gamma_{q\ j}^{\phantom\ i }$, $\Gamma_{0\ i}^{\phantom\ 0 }$ and
$\Gamma_{i\ j}^{\phantom\ 0 }$.

The evolution equation for $\Gamma_{q\ j}^{\phantom\ i }$ is found  using equation \il(\ref{E:D41}), where
\be\label{E:rue1}
\partial_t \Gamma_{q\ j}^{\phantom\ i }=R^i{}_{j0q}-\Gamma_{k\ j}^{\phantom\ i }\Gamma_{q\ 0}^{\phantom\ k}-\Gamma_{0\ 0}^{\phantom\ i }\Gamma_{q\ j}^{\phantom\ 0}+\Gamma_{0\ j}^{\phantom\ 0 }\Gamma_{q\ 0}^{\phantom\ i }.
\ee

Now, using equation \il(\ref{E:recie}) one can express the component of the Riemann tensor $R^i{}_{j0q}$ in terms of the matter fields. Using the Einstein equation $R_{0p}=\kappa T_{0p}$, with $T_{0p}=T^{\ti{(em)}}_{0p}+T^{\ti{(KG)}}_{0p}$ and
\be\label{E:berp}
T^{\ti{(em)}}_{0p}\equiv-g^{ab}F_{0a}F_{b p},
\qquad
\Tkg_{0p}=T^{(\phi)}_{0p}+\frac{\phi^2}{2}T^{(\sigma)}_{0p}+\frac{\phi^2}{2}T^{\ti{(A)}}_{0p}+
\frac{\phi^2}{2}T^{(\sigma \ti{A})}_{0p},
\ee
where
\bea\label{E:berpfin}
\hspace{-0.7cm}T^{(\phi)}_{0p}&=& \partial_0\phi\partial_p\phi=0,
\quad
T^{(\sigma)}_{0p}=
2\sigma_0\sigma_p,
\quad
T^{\ti{(A)}}_{0p}= 2
A_0A_p,
\quad
T^{(\sigma \ti{A})}_{0p}= 4
\sigma_{(0}A_{p)},
\eea
we have
\be\label{E:ricciop2}
R_{0p}=\kappa\bigg(-g^{ab}F_{0a}F_{b p}+\phi^2\big(\sigma_0\sigma_p+
A_0A_p+2
\sigma_{(0}A_{p)}\big)\bigg).
\ee
Accordingly, the component $R_{0p}$ can be expressed purely in terms
of undifferentiated field variables.

\medskip
The evolution equations for the coefficients $\Gamma_{0\ i}^{\phantom\
0 }$ and $\Gamma_{q\ j}^{\phantom\ 0 }$ can be found again using
equation \il(\ref{E:D41}). One finds that
\be\label{E:rue2}
\partial_t \Gamma_{q\ j}^{\phantom\ 0}=e_q \Gamma_{0\ j}^{\phantom\ 0}+R^{0}{}_{j0q}+ \Gamma_{0\ j}^{\phantom\ 0} \Gamma_{0\ q}^{\phantom\ 0}-\Gamma_{i\ j}^{\phantom\ 0}\Gamma_{q\ 0}^{\phantom\ i}-\Gamma_{0\ i}^{\phantom\ 0}\Gamma_{q\ j}^{\phantom\ i}.
\ee
Now, using the conservation equations  and the
condition $\nabla_{[a}\nabla_{b]}\phi=0$ one has the symmetry $\g{0}_{i\ j}=\g{0}_{j\ i}$. From the Klein Gordon equation for the field $\phi$ we can recover the following evolution equation for the variable $\alpha$:
\be\label{E:alphaew}
\dot{\alpha}-\alpha \chi-\phi(\sigma^2+ A^2+2 \sigma^aA_a)=0.
\ee
Now,  the evolution equation for $\Gamma_{0\ i}^{\phantom\ 0 }$  can be found observing that
\be\label{E:congiu}
[e_0,e_i]\alpha=\Gamma_{0\ i}^{\phantom\ 0 }\partial_t\alpha-\chi_i^{\phantom\ j}\nabla_j\alpha.
\ee
We can use equation \il(\ref{E:alphaew}) for the term  $\partial_t\alpha$. From the commutator $[e_i,e_0]\phi=0$ we obtain the constraint
\be
\nabla_i\alpha-\g{0}_{0\ i}\alpha=0.
\ee
Therefore equation \il(\ref{E:congiu}) can be rewritten as
\be
[e_0,e_i]\alpha=\Gamma_{0\ i}^{\phantom\ 0 }\alpha \chi-\chi_i^{\phantom\ j}(\g{0}_{0\ j}\alpha)+\Gamma_{0\ i}^{\phantom\ 0 }\phi(\sigma^2+ A^2+2q \sigma^aA_a),
\ee
so as to obtain
\be\label{E:trova}
-(\nabla_0 \g{0}_{0\ i}+ \nabla_i\chi)=\g{0}_{0\ i} \chi-\chi_i^{\phantom\ j}\g{0}_{0\ j}+\frac{2}{\alpha}\g{0}_{0 \ i}\frac{d\bar{\nu}}{d\phi}+\frac{2}{\alpha}\nabla_i\frac{d\bar{\nu}}{d\phi},
\ee
where
\be
\frac{d\bar{\nu}}{d\phi}\equiv\phi(\sigma^2+ A^2+2 \sigma^aA_a).
\ee
The last term of equation \il(\ref{E:trova}) contains
$\nabla_i\sigma^a$ and $\nabla_iA_a$. However, making use of the
auxiliary variables $\psi_{ab}$ and $\zeta_{ab}$, it can be expressed
in terms of undifferentiated quantities. Thus,
noting that the component $R_{0p}$ of the Ricci tensor is
\be\label{E:ricciop}
R_{0p}=-[e_p \Gamma_{i\ 0}^{\phantom\ i}-e_i \Gamma_{p\ 0}^{\phantom\ i}+ \Gamma_{j\ 0}^{\phantom\ i}( \Gamma_{p\ i}^{\phantom\ j}-\Gamma_{i\ p}^{\phantom\ j}) +\Gamma_{j\ 0}^{\phantom\ i}\Gamma_{i\ p}^{\phantom\ j}-\Gamma_{i\ j}^{\phantom\ i}\Gamma_{p\ 0}^{\phantom\ j}],
\ee
and using equation \il(\ref{E:ricciop}) in equation \il(\ref{E:ricciop2}) it follows
\bea\nonumber
-[e_p \Gamma_{i\ 0}^{\phantom\ i}-e_i \Gamma_{p\ 0}^{\phantom\ i}+ \Gamma_{j\ 0}^{\phantom\ i}( \Gamma_{p\ i}^{\phantom\ j}-\Gamma_{i\ p}^{\phantom\ j})+\Gamma_{j\ 0}^{\phantom\ i}\Gamma_{i\ p}^{\phantom\ j}-\Gamma_{i\ j}^{\phantom\ i}\Gamma_{p\ 0}^{\phantom\ j}]
=
\\\label{E:teg}
\kappa\left[-g^{ab}F_{0a}F_{b p}+\phi^2\left(\sigma_0\sigma_p+
A_0A_p+2
\sigma_{(0}A_{p)}\right)\right].\eea
Substituting in equation \il(\ref{E:trova}), the term $(-
\nabla_i\chi)$, where $\chi=- h^{cd}\Gamma_{cd}^0$, and noting that
$-e_p \Gamma_{i\ 0}^{\phantom\ i}=e_ph^{ij}\g{0}_{i\ j}$ we have, from
equation \il(\ref{E:teg})
\bea
\nabla_p\chi=-e_i \Gamma_{p\ 0}^{\phantom\ i}+ \Gamma_{j\ 0}^{\phantom\ i}( \Gamma_{p\ i}^{\phantom\ j}-\Gamma_{i\ p}^{\phantom\ j})+\Gamma_{j\ 0}^{\phantom\ i}\Gamma_{i\ p}^{\phantom\ j}-\Gamma_{i\ j}^{\phantom\ i}\Gamma_{p\ 0}^{\phantom\ j}+
\\
\kappa\left[-g^{ab}F_{0a}F_{b p}+\phi^2\left(\sigma_0\sigma_p+
A_0A_p+2
\sigma_{(0}A_{p)}\right)\right],
\eea
thus equation \il(\ref{E:trova}) reads
\bea\label{E:trovag}
&& \nabla_0 \g{0}_{0\ i}- e_p \Gamma_{i\ 0}^{\phantom\ p}=-(\g{0}_{0\ i} \chi-\chi_i^{\phantom\ j}\g{0}_{0\ j}+\frac{2}{\alpha}\g{0}_{0 \ i}\frac{d\bar{\nu}}{d\phi}+\frac{2}{\alpha}\nabla_i\frac{d\bar{\nu}}{d\phi})
\\
&& \nonumber-\left\{\Gamma_{j\ 0}^{\phantom\ p}\Gamma_{p\ i}^{\phantom\ j}-\Gamma_{p\ j}^{\phantom\ p}\Gamma_{i\ 0}^{\phantom\ j}+\kappa\left[-g^{ab}F_{0a}F_{b i}+\phi^2\left(\sigma_0\sigma_i+
A_0A_i+2
\sigma_{(0}A_{i)}\right)\right]\right\},
\eea
where we used  $\g{0}_{i\ j}=\g{0}_{j\ i}$. Equation (\ref{E:trovag}) is therefore the evolution equation for the  connection $\g{0}_{0\ i}$.

\medskip
 We concentrate now our analysis on equation \il(\ref{E:rue2}) for the
case of the components $\Gamma_{q\ j}^{\phantom\ 0}$. Using equation
\il(\ref{E:decom}), the curvature component $R^{0}_{j0q}$ can be split
as
\be\label{E:decomf}
R^0_{j0q}=C^0{}_{j0q}+ \frac{1}{2}\left(S_{qj}+g_{jq}S_{0}^{\phantom \ 0}\right).
\ee
Thus, one finds that
\be\label{E:rue211}
\partial_t \Gamma_{q\ j}^{\phantom\ 0}-e_q \Gamma_{0\ j}^{\phantom\ 0}=C^0{}_{j0q}+ \frac{1}{2}\left(S_{qj}+g_{jq}S_{0}^{\phantom \ 0}\right) + \Gamma_{0\ j}^{\phantom\ 0} \Gamma_{0\ q}^{\phantom\ 0}-\Gamma_{i\ j}^{\phantom\ 0}\Gamma_{q\ 0}^{\phantom\ i}-\Gamma_{0\ i}^{\phantom\ 0}\Gamma_{q\ j}^{\phantom\ i},
\ee
where
\be\label{E:sabqw}
 S_{ab}=S^{\ti{(KG)}}_{ab}+S^{\ti{(em)}}_{ab},
\qquad S_{ab}\equiv \kappa(T_{ab}-\frac{1}{3}g_{ab}g^{cd} T_{cd}).
\ee
Using equations \il(\ref{E:aaa}), \il(\ref{E:extendee}) and
\il(\ref{E:berp})--(\ref{E:berpfin}), one can readily conclude that that the right hand side of equation \il(\ref{E:rue211}) does
not contain derivatives of the field unknowns.

\medskip
Summarizing, equations (\ref{E:trovag}) and (\ref{E:rue211}) provide
the required evolution equations for the connection components
$\g{0}_{0\ i}$ and $\Gamma_{q\ j}^{\phantom\ 0}$, respectively.

\subsection{Summary of the analysis}
\label{Section:Summary}
We now summarizes the analysis of the evolution equations carried out
in the preceding sections. The unknowns of the evolution equations
can be conveniently grouped in the vector
\be\label{E:v} \textbf{{v}}=\big(e_i{}^\mu,\Gamma_{0\ i}^{\phantom\ 0 },\Gamma_{q\ j}^{\phantom\ i },
 \Gamma_{q\ j}^{\phantom\ 0 },\hat{E}_{ab},\hat{B}_{ab},E_{a},B_{a},\mathcal{E}_{a},
\mathcal{B}_{a},\phi,A_a,\sigma_a,
\alpha,
\psi_{ab},
\zeta_{ab}\big).
\ee
The components of $\Gamma_a{}^b{}_c$ not
included in this list are determined by means of gauge conditions and
symmetries. By construction, the electric and magnetic parts of the
Weyl tensor are tracefree. This symmetry is disregarded in these
considerations and is recovered by imposing it on the initial data. It
can be shown that if these tensors are initially tracefree, then they
will be also tracefree for all later times ---see
e.g. \cite{FriRen00}.

\medskip
The evolution equations for the independent components of the unknowns
in (\ref{E:v}) and the underlying assumptions in their construction
are given as follows:

\begin{itemize}
\item[(i)] The evolution equation for the tetrad coefficients,
$e_i{}^\mu$, is given by equation \il(\ref{E:h2o1oboe}) by virtue of the
Lagrange condition $N^a=e^a_0=\delta^a_0$. Equation \il(\ref{E:h2o1oboe})
has the same principal part than the corresponding equation in the
case of an uncharged perfect fluid ---see e.g. \cite{Fri98c}. It gives rise
to a symmetric hyperbolic subsystem of equations.

\item[(ii)] The the evolution equations for the connection
coefficients, $\g{0}_{0\ i}$ and $\g{a}_{i\ b}$ are given,
respectively, by equations \il(\ref{E:trovag}) and
\il(\ref{E:rue211}). As a consequence of the gauge conditions one has,
in addition, that $\Gamma_{0\i}^{\phantom\ j }=0$.  Equation
\il(\ref{E:rue1}) also takes care of $\g{q}_{j\ k}$. As in the case of
the equations frame coefficients, equations \il(\ref{E:trovag}),
\il(\ref{E:rue211}) and \il(\ref{E:rue1}) have the same principal part
as those of the analysis given in \cite{Fri98c}. Again, one has an
hyperbolic subsystem of equations.

\item[(iii)] The evolution equations for the electric and magnetic
parts of the Weyl tensor, $\hat{E}_{ab}$ and $\hat{B}_{ab}$, are
given, respectively, by equations (\ref{E:terete}) and
\il(\ref{E:sa}). As mentioned before, the tracefreeness of these
tensors is not used to reduce the number of independent
components. Thus one has 12 equations for equal number of
components. Equations with a principal part of the form of
equations (\ref{E:terete}) and \il(\ref{E:sa}) are symmetric hyperbolic
independently of the gauge choice ---see e.g. \cite{FriRen00}.

\item[(iv)] The evolution equation for the electric and magnetic parts
of the Faraday tensor, $E_a$ and $B_b$, are given, respectively, by
equations \il(\ref{E:NE2a}) and \il(\ref{E:Nb2a}).  As in the case of
the equations for the electric and magnetic parts of the Weyl tensor,
the principal part of these equations is known to be hyperbolic
independently of the gauge ---again, see e.g. \cite{FriRen00}.

\item[(v)] The evolution equations for the electric and magnetic
parts of parts of the auxiliary field $\psi_{abc}$ (encoding the
covariant derivative of the Faraday tensor), $\mathcal{E}_a$ and
$\mathcal{B}_b$ are given, respectively, by equations
\il(\ref{E:NE2ab}) and \il(\ref{E:Nb2ab}).  These equations involve 24
equations for as many unknowns. Their structure is analogous to that
of equations \il(\ref{E:NE2a}) and \il(\ref{E:Nb2a}), except that
they contain an extra free index. As a consequence, their principal
part gives rise to a symmetric hyperbolic subsystem.

\item[(vi)] The evolution equation for the vector potential $A_a$ is
given by equations \il(\ref{E:Aeq}) and \il(\ref{E:Aeqt}). The
symmetric hyperbolicity of the equations is implemented by the
addition of the (arbitrary) gauge source function $\zeta(x)$ which
allows to freely specify the divergence $\nabla_a A^a$.

\item[(vii)] The evolution equation for the scalar field $\phi$, is
given by equation \il(\ref{E:fie}).

\item[(viii)] The evolution equations for the auxiliary field
$\sigma_a\equiv\nabla_a \theta$ are given by equations
\il(\ref{E:sigma}) and \il(\ref{E:sigmat}) with the understanding that
the derivative $\nabla_a\sigma^a=\psi(x)$ arising in the right-hand
sides is recovered from the Klein Gordon equation in the form given by
\il(\ref{E:forsinowma}).

\item[(ix)] The equation for the scalar $\alpha$ defined by the
conditions $\alpha^2 \equiv \varphi$ and
$\varphi^2=\nabla_a\phi\nabla^a\phi$ is given by equation
\il(\ref{E:alphaew}). This evolution equation is a consequence of the
Klein Gordon equation for the field $\phi$ ---see equation
\il(\ref{E:forsifi}).

\item[(x)] The evolution equations for the components of $\psi_{ab}\equiv\nabla_a\psi_b$ are given by equations \il(\ref{E:psi}) and  (\ref{E:psit}).

\item[(xi)] Finally, the evolution equations for the components of
$\zeta_{ab}\equiv\nabla_aA_b$ are given by equations
\il(\ref{E:stasst}) and (\ref{E:stasstt}).

\end{itemize}

\medskip
Summarizing, equations. (\ref{E:h2o1oboe}), \il(\ref{E:trovag}),
\il(\ref{E:rue211}), (\ref{E:rue1}), (\ref{E:terete}), (\ref{E:sa}),
(\ref{E:NE2a}), (\ref{E:Nb2a}), (\ref{E:NE2ab}), (\ref{E:Nb2ab}),
(\ref{E:Aeq}), (\ref{E:Aeqt}), (\ref{E:sigma}), (\ref{E:sigmat}),
(\ref{E:alphaew}), (\ref{E:psi}), (\ref{E:psit}), (\ref{E:stasst}) and
(\ref{E:stasstt}), provide the desired symmetric hyperbolic evolution
system for the Einstein-charged scalar field system. This system can
be written in a concise form using matrix notation. One has that
\begin{equation}
\label{E:II}
\mathbf{A}^0\partial_0 \textbf{{v}}-\mathbf{A}^j\partial_j \textbf{{v}}=\mathbf{B} \mathbf{v},
\end{equation}
where $\mathbf{A}^\mu=\mathbf{A}^0(x^{\mu},\mathbf{v})$ are
matrix-valued functions of the coordinates and the unknowns
$\mathbf{v}$, and $\mathbf{A}^{0}$ is positive definite at least close
to a fiducial initial hypersurface. The structure of the system
(\ref{E:II}) ensures the existence of local solutions to the evolution
equations. The analysis of whether these solutions give rise to a
solution of the full Einstein-charged scalar field system requires the
analysis of the evolution of the constraint equations. This is a
computationally intensive argument which will be omitted here. A
general argument to handle the propagation of the constraints without
having to resort to lengthy computations can be found in \cite{Reu98}.

\section{Conclusion}\label{Sec:Colcusione}
In the present article we have revisited the issue of well-posedness
initial value problem for the evolution equations of the
Einstein-Maxwell-Klein-Gordon  system (a self-gravitating charged scalar field).  The
approach followed makes use of the well known $1+3$ tetrad formalism
by means of which the various tensorial quantities and equations are
projected along the direction of the comoving observer and onto the
orthogonal subspace. Following \cite{Fri98c,FriRen00}, we require the
timelike vector of the orthonormal frame to follow the matter flow
lines (Lagrangian gauge). {The  adapted frame  for this problem has been constructed taking $e_0=\varphi_a/\alpha$. The gauge condition depends therefore on the scalar field evolution, the construction  of the  appropriate frame for this problem  relies on   suitable regularity  conditions for $\phi$. As discussed in Section\il\ref{Sec:NVGC}
it is assumed  that
$\phi\in \mathcal{C}^{\infty}(\mathcal{M})$ and  $\nabla_a\phi$
is timelike.}  Moreover, we assume the vector fields
tetrad to be Fermi transported in the direction of $U$, these
conditions fix certain components of the connection.

\medskip
 A key feature of our analysis was the introduction of several auxiliary fields, the
 tensors $\psi_{abc},\zeta_{ab}, \psi_{ab}$ corresponding to the covariant
derivative of the Faraday tensor $F_{ab}$, the vector potential $A_a$ and the scalar field $\phi$. The purpose of introducing
these tensors was to ensure the symmetric hyperbolicity of the
evolution equations for the components of the Weyl tensor.

\section*{Acknowledgments}
DP gratefully acknowledges financial support from the A. Della Riccia
Foundation and  Blanceflor Boncompagni-Ludovisi n\'ee Bildt Foundation.
\addcontentsline{toc}{chapter}{Bibliography}


\end{document}